\documentclass{IEEEtran}
\usepackage{url}
\usepackage{setspace}
\usepackage{cite}
\usepackage{amsmath,amssymb,amsfonts}
\usepackage{graphicx}
\usepackage{textcomp,nicefrac,comment}
\usepackage[switch]{lineno}
\def\BibTeX{{\rm B\kern-.05em{\sc i\kern-.025em b}\kern-.08emT\kern-.1667em\lower.7ex\hbox{E}\kern-.125emX}}
\begin{document}
\IEEEoverridecommandlockouts
\IEEEpubid{
  \parbox{\columnwidth}{
    %\copyright\space Copyright 2020 CERN for the benefit of the ATLAS Collaboration. CC-BY-4.0 license
    From ATL-DAQ-PROC-2020-021. Published with permission by CERN.
    \hfill}
  \hspace{0.9\columnsep}\makebox[\columnwidth]{\hfill}}
\IEEEpubidadjcol

\makeatletter
\def\@IEEEpubidpullup{2\baselineskip}
\makeatother
\title{ATLAS hardware-based Endcap Muon Trigger \\ for future upgrades}
\author{Yuya Mino$^\ast$, on behalf of ATLAS Collaboration \\
  $^\ast$Kyoto University Graduate School of Science, Kyoto, Japan}

%new commands
\newcommand{\ms}{$\mu\mathrm{s}$}
\newcommand{\mum}{$\mu\mathrm{m}$}
\newcommand{\lumi}{$\times10^{34} \rm \ cm^{-2} s^{-1}$ }
\newcommand{\fb}{$\rm fb^{-1}$}
\newcommand{\pT}{$p_{\rm{T}}$}
\newcommand{\ET}{$E_{\rm T}$}
\newcommand{\MET}{$E_{\rm T}^{\rm miss}$}

%\footnote{Copyright 2020 CERN for the benefit of the ATLAS Collaboration. CC-BY-4.0 license}

\maketitle

%\linenumbers

\begin{abstract}
  The LHC is expected to increase its center-of-mass energy from 13 TeV to 14 TeV for \mbox{Run 3} scheduled from 2022 to 2024.
  After \mbox{Run 3}, upgrades for the High-Luminosity-LHC (\mbox{HL-LHC}) programme are planned and the operation will start in 2027, increasing the instantaneous luminosity to 5.0 -- 7.5 times its nominal luminosity.
  Continuous upgrades of the ATLAS trigger system are planned to cope with the high event rate and to keep the physics acceptance.
  During the long shutdown period before \mbox{Run 3}, new detectors will be installed to improve the trigger performance.
  New trigger logic, combining information from detectors located outside of the magnetic field and new detectors installed inside the magnetic field, are introduced from \mbox{Run 3} to reduce the trigger rate.
  In order to handle data from the various detectors, a new trigger processor board has been developed and the design is presented.
  During the upgrade for \mbox{HL-LHC}, the trigger and readout systems of the first level hardware-based part are planned to be upgraded.
  Full-granularity information will be transferred to the trigger processor board which enables more off-line like track reconstruction in the hardware-based system.
  To handle the full-granularity information and perform the hardware-based track reconstruction, the trigger processor board will implement an FPGA with hundreds of pairs of transceivers and huge memory resources.
  Expected performance for the hardware-based endcap muon trigger in \mbox{Run 3} and \mbox{HL-LHC}  will also be presented.
\end{abstract}


\section{Introduction}
\label{sec:introduction}
\IEEEPARstart{T}{he} Standard Model is a theory which provides the best explanation of elementary particles and their interactions for three of the four known fundamental forces (electromagnetic, weak and strong interactions).
Although the Standard Model has succeeded in explaining almost all experimental results, there are some problems which cannot be explained, such as the hierarchy problem and the origin of dark matter.
Experimental verification of new physics (Beyond Standard Model, BSM) is required.

\begin{figure}[tb]
\centerline{\includegraphics[width=3.5in]{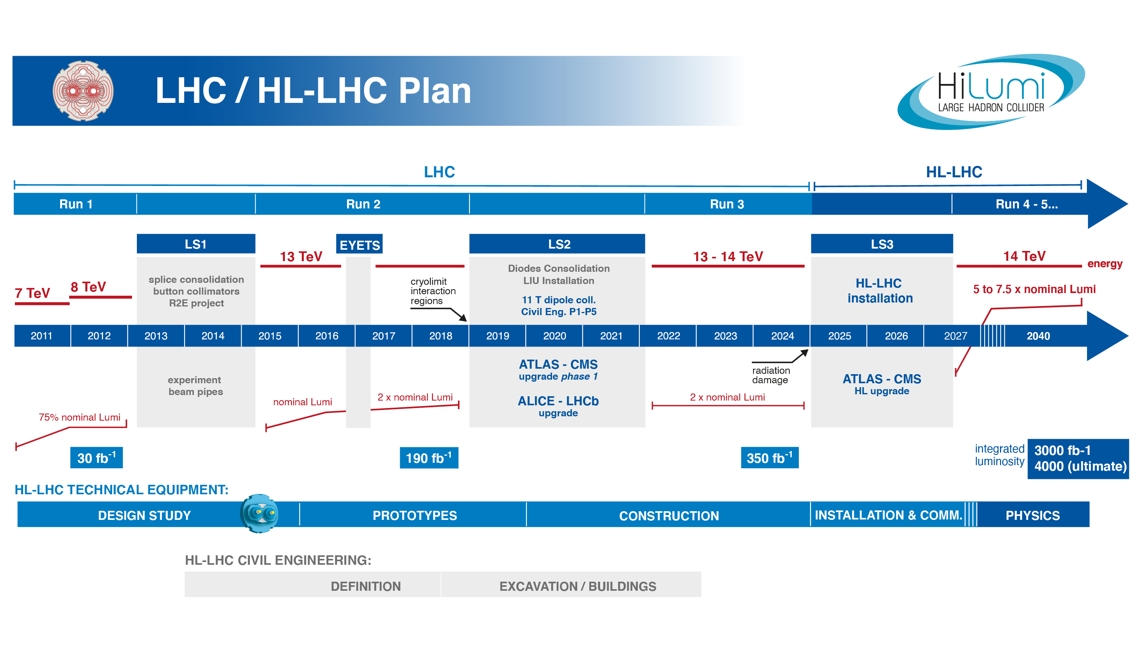}}
\caption{The LHC and \mbox{HL-LHC} upgrade plans\cite{schedule}. \mbox{Run 3} will start in 2022 and increase the center-of-mass energy to $\sqrt{s}$ = 14 TeV during the run. \mbox{HL-LHC} will start in 2027 and increase the luminosity to 5.0 -- 7.5 times the nominal luminosity.}
\label{schedule}
\end{figure}

The Large Hadron Collider (LHC) \cite{LHC} is the world's largest accelerator, colliding protons with center-of-mass energy of $\sqrt{s}$ = 13 TeV and peak instantaneous luminosity of \mbox{2.0 \lumi}.
Future upgrades are planned for BSM searches and for Standard Model precision studies with higher energy and luminosity as shown in \figurename~\ref{schedule}.
\mbox{Run 3} is planned to raise the center-of-mass energy from 13 TeV to 14 TeV and collect data from 2022 to 2024 to extend the parameter space for various new physics models.
The \mbox{HL-LHC} is planned to increase the peak instantaneous luminosity to \mbox{7.5 \lumi} and collect data for ten years, which corresponds to a total integrated luminosity of 3000 \fb.

%LHC の説明と schedule

%ATLAS 検出器の説明
The ATLAS detector \cite{ATLAS} is a general-purpose detector at the LHC, investigating a wide range of physics using data from proton-proton collisions at 40 MHz frequency.
Since the final recording rate of data from physics events is limited to approximately 1 kHz on average, ATLAS uses a two-level online trigger system to select events from interesting physics processes.
The ATLAS trigger system consists of a hardware-based Level-1 (L1) trigger and a software-based High-Level trigger (HLT).
The L1 trigger uses a subset of information from the detector to make decisions and reduces the event rate to 100 kHz.
The decision is made in 2.5 \ms~(called the L1 latency).
The HLT trigger receives event candidates from the L1 trigger and refines the decision using the full detector information.

%Phase-I upgrade について
The ATLAS detector will include new L1 trigger capabilities for \mbox{Run 3}.
However, the limitation of the L1 trigger rate will be kept at 100 kHz. 
New detectors will be installed to improve performance and reduce the trigger rate of current triggers.
This period is called the Phase-I Upgrade.

%Phase-II upgrade について
In order to cope with the higher luminosity in the \mbox{HL-LHC}, the trigger and readout systems of the hardware-based trigger are planned to be further upgraded.
The trigger latency and rate will be increased to 10 \ms~and 1 MHz, respectively, by replacing the current system with high-bandwidth readout electronics.
The increased latency will enable much more sophisticated algorithms to improve the trigger performance.
Along with this upgrade, the L1 trigger will be renamed to Level-0 (L0) trigger after \mbox{Run 3}.
This period is called the Phase-II Upgrade.

%%%%%%%%%%%%%%%%%%%%%%%%%%%%%%%%%%%%%%%%%%%%%%%
%% \mbox{Run 3}                                     %%
%%%%%%%%%%%%%%%%%%%%%%%%%%%%%%%%%%%%%%%%%%%%%%%
%\newpage
\section{PHASE-I UPGRADE OF \\ THE ATLAS LEVEL-1 MUON TRIGGER}
%Run 2 muon trigger とその問題点について (pT の測り方から)
The rate of the single muon trigger with the highest \pT~threshold (``primary muon trigger'') in \mbox{Run 3} is required to be reduced to 15 kHz considering other trigger and physics requirements \cite{Phase1TDR}.
With the current trigger scheme in \mbox{Run 2}, the trigger rate could not be reduced below 20 kHz at an instantaneous luminosity of \mbox{2.0 \lumi}.
Raising the \pT~threshold would reduce the trigger rate but would also reduce the physics acceptance.
%muon trigger の重要性について
%Higgs production is one of the most important processes measured in the ATLAS experiment.
For example, Higgs-strahlung from $W$ bosons is one of the production processes to determine the Higgs boson coupling to the gauge bosons and fermions at high precision.
Leptons from the $W$ boson are used to trigger this process and 93\% of the muons have \pT~larger than \mbox{20 GeV}.
If the \pT~threshold is raised to 40 GeV, more than 30\% of events from the Higgs-strahlung process would be lost \cite{NSW}.
Thus, the upgrade of the current trigger scheme is mandatory to keep the physics acceptance.

%\begin{figure}[b]
%\centerline{\includegraphics[width=2.5in]{fig/Acceptance.pdf}}
%\caption{
%  \pT~distribution of muons from $W$ boson in the Higgs-strahlung process\cite{NSW}.
%  Efficiency of the muon trigger decreases from 93\% to 61\% if the \pT~threshold is raised from \mbox{20 GeV} to 40 GeV.
%}
%\label{Acceptance}
%\end{figure}

\begin{figure}[tbh]
\centerline{\includegraphics[width=3.0in]{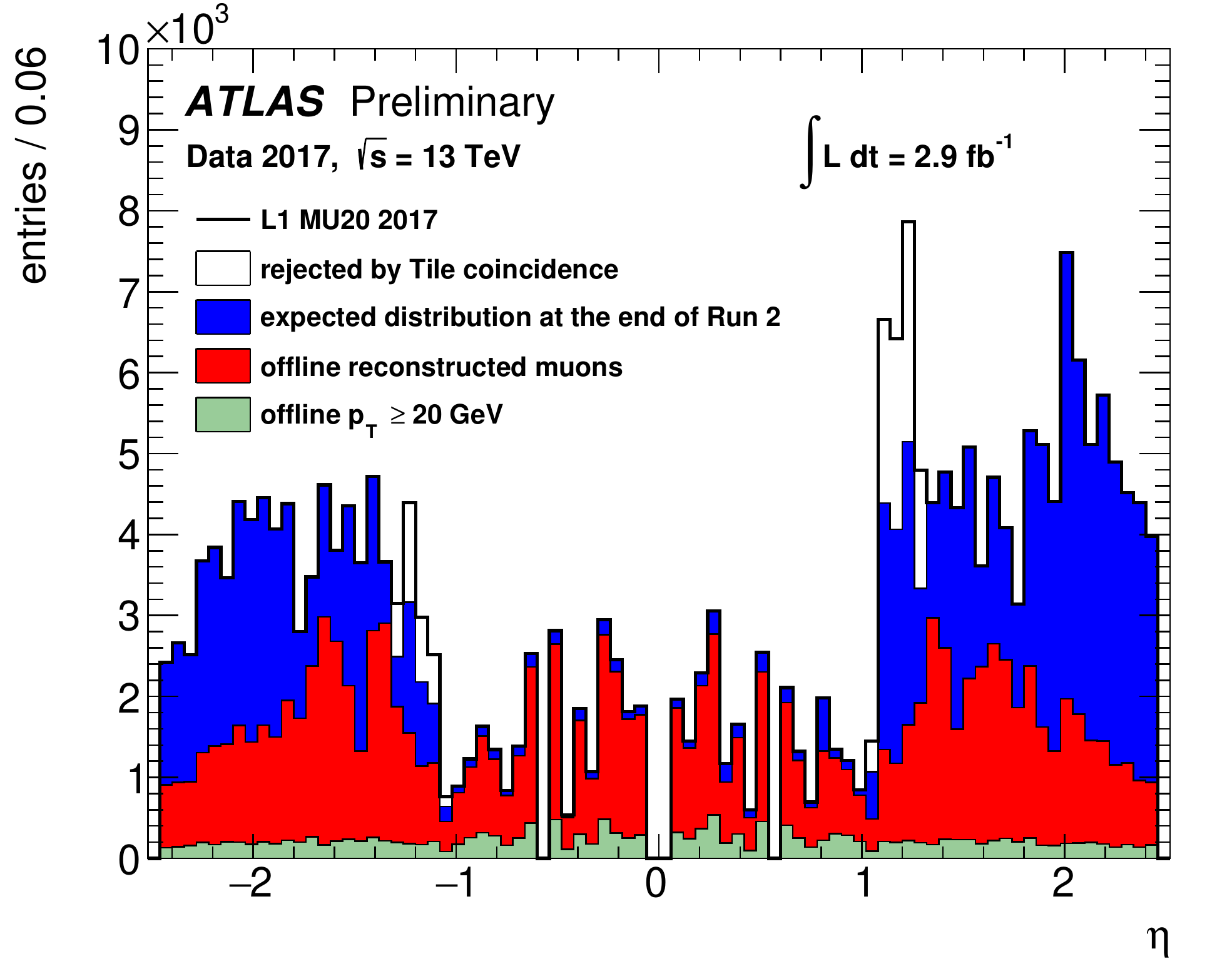}}
\caption{
  $\eta$ distribution of trigger candidates from L1\_MU20 \cite{L1Public}.
  Blue and red regions show track candidates from fake and low \pT~muons, respectively.
}
\label{Run2_Trigger}
\end{figure}

%Run 2 trigger について
During \mbox{Run 2}, the L1 muon trigger rate was dominated by low \pT~muons below the \pT~threshold and charged particles emerging from the endcap toroidal magnets (``fake'' muons).
Figure~\ref{Run2_Trigger} shows the $\eta$ distribution of trigger candidates in the L1 single muon trigger at \pT~\mbox{20 GeV} in \mbox{Run 2} (L1\_MU20).
Charged particles from the endcap toroidal magnets are bent by the magnetic field, so only the positive-charged particles point in the direction of the interaction point, for the A-side.
As shown in \figurename~\ref{Run2_Trigger}, the number of track candidates from fake muons is larger in the A-side compared to the C-side because most of the fake muons originate from protons with positive charge.

Approximately 80 \% of the trigger candidates are from the endcap region ($|\eta| > 1.05$).
In order to reduce triggers from low \pT~and fake muons in the endcap region, new algorithms are implemented using information from the new detectors installed for \mbox{Run 3}.

\begin{figure}[tbh]
\centerline{\includegraphics[width=3.5in]{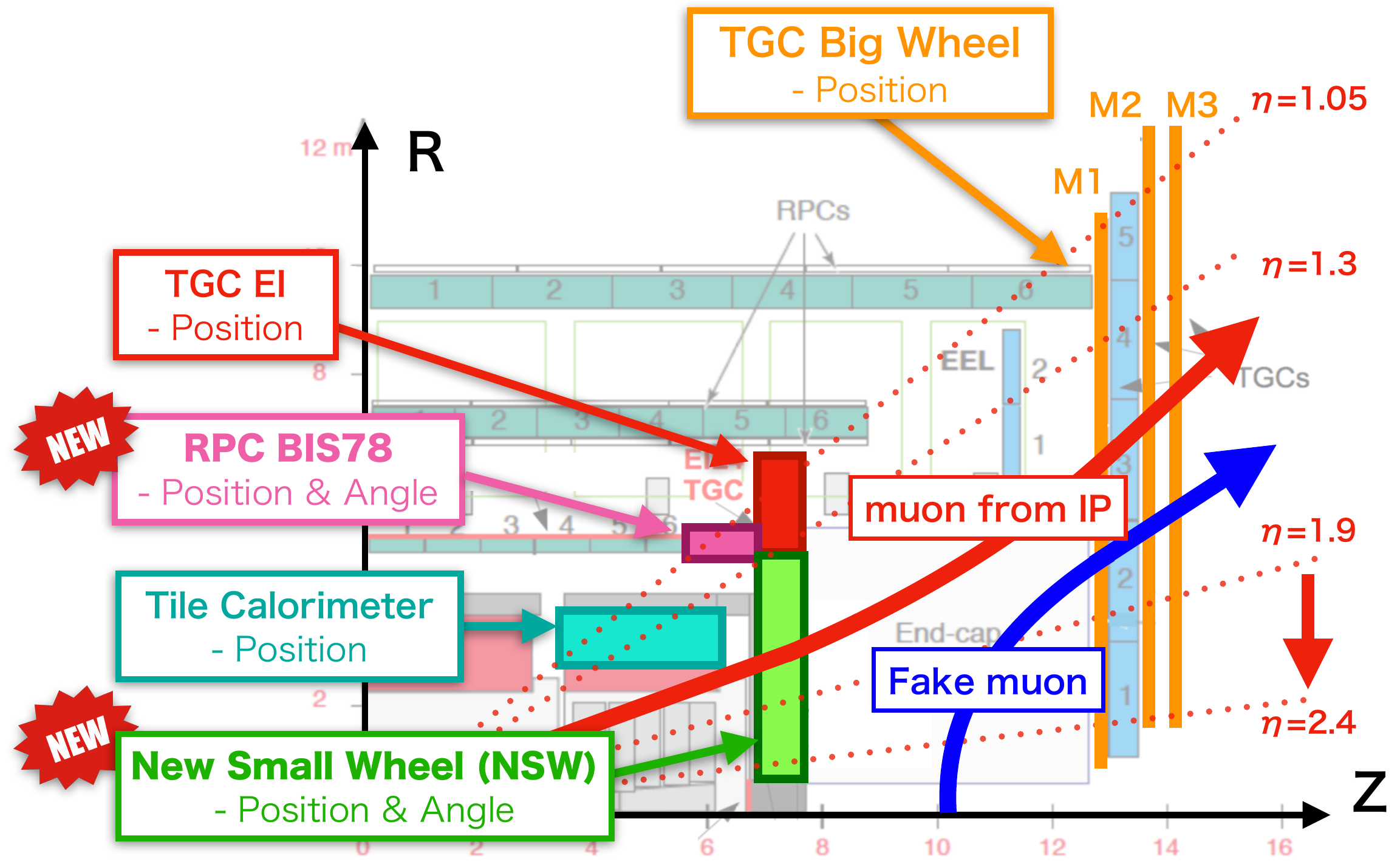}}
\caption{Layout of the ATLAS muon spectrometer in \mbox{Run 3}. In the current system, detector coverage in the magnetic field is limited to $|\eta| = 1.9$. By installing the NSW, the detector coverage will be extended to $|\eta| = 2.4$.}
\label{Run3_Detector}
\end{figure}

Figure~\ref{Run3_Detector} shows the detectors used in the L1 endcap muon trigger from \mbox{Run 3}.
Muons passing the toroidal magnetic field are bent and their position information measured by Thin-Gap Chambers (TGC) \cite{TGC} is used to measure \pT.
The TGC chambers are aligned in a disk-shaped structure called the TGC Big Wheel (TGC-BW) and the BWs are placed on both sides of the ATLAS detector.
%fake muon の除去方法について
The fake muons leave hits in the TGC-BW which imitate hits from high \pT~muons from the interaction point, as shown in \figurename~\ref{Run3_Detector}.
However, the fake muons do not leave hits in the detectors inside the magnetic field since they emerge directly from the toroidal magnet.
Thus, the fake muons are reduced by combining hit information from the TGC-BW and detectors inside the magnetic field. 
Various detectors are placed inside the magnetic field and used in the L1 endcap muon trigger in \mbox{Run 3}: New Small Wheel (NSW) \cite{NSW}, TGC in the endcap inner station (TGC EI) \cite{TGC}, Resistive Plate Chambers in the barrel inner station (RPC BIS78) \cite{RPCBIS78} and Tile hadronic calorimeter (TileCal) \cite{TileCal}.
The NSW and RPCBIS78 are new detectors installed during the Phase-I Upgrade. 
The NSW consists of 8 layers each of sTGC (small-strip TGC) and micromegas \cite{Micromegas}.
Combining hit information from the multiple layers enables reconstruction of tracks with angular resolution of 1 mrad in the $\theta$ direction.
Position resolution of the NSW is 0.005 and 10 mrad in the $\eta$ and $\phi$ directions, which is much better than the current detector replaced by the NSW, which measures position with resolution of 0.15 and 65 mrad in the $\eta$ and $\phi$ directions.
Coverage of detectors inside the magnetic field will be extended from $|\eta| = 1.9$ to $|\eta| = 2.4$, which results in further reduction of the fake muons.

\subsection{New coincidence logic in \mbox{Run 3}}
The new trigger algorithms using new detectors in the L1 endcap muon trigger system in \mbox{Run 3} are introduced in this section.

\vspace{\baselineskip}
\subsubsection{Position matching}
\begin{figure}[tbh]
\centerline{\includegraphics[width=3.5in]{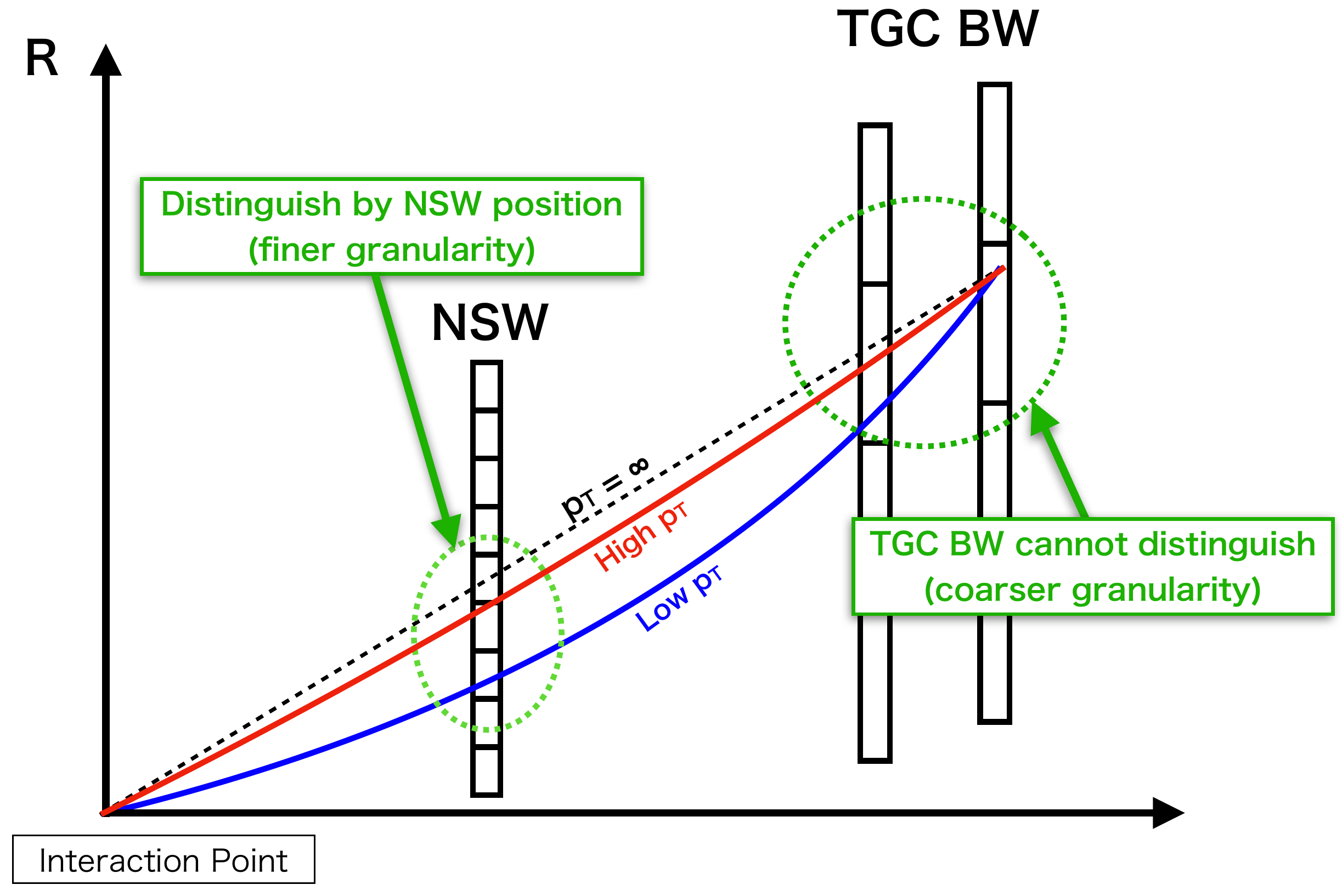}}
\caption{
  Schematic of position matching algorithm.
  Low \pT~and high \pT~muons can be distinguished by using postion information of the new detectors with finer granularity. 
}
\label{PositionMatching}
\end{figure}

The \pT~resolution of a track candidate found by the TGC-BW is limited by the detector granularity as shown in \figurename~\ref{PositionMatching}. 
The \pT~resolution can be improved by refining the \pT~with position difference between the TGC-BW and new detectors in the $\eta$ and $\phi$ directions because new detectors have finer granularity.
By requiring the position difference to have an appropriate value, the low \pT~muons which could not be reduced by the hit information in the TGC-BW are reduced.
The position difference information is handed over to a \mbox{Look-Up-Table (LUT)} implemented on a Field Programmable Gate Array (FPGA) and the corresponding \pT~is immediately returned to refine the \pT~decision of the TGC-BW.
LUTs are defined depending on the position of the trigger candidate since the toroidal magnetic field is non-uniform and the correlation between \pT~and position difference differs.

\vspace{\baselineskip}
\subsubsection{Angle matching}
\begin{figure}[tbh]
\centerline{\includegraphics[width=3.5in]{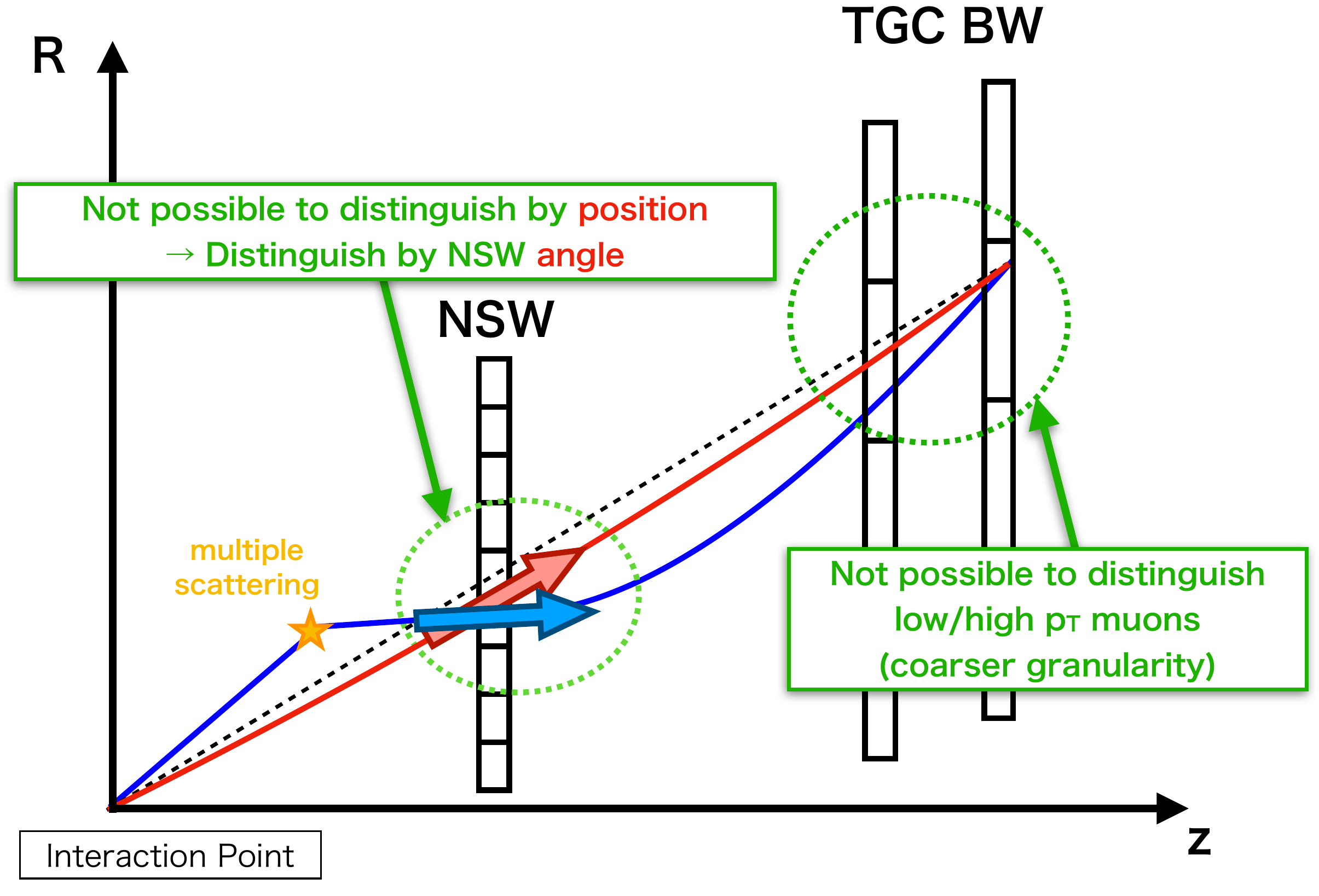}}
\caption{
  Schematic of angle matching algorithm.
  Combination of the angle information and position difference enables further reduction of the remaining low \pT~muons from the beam spot size and multiple scatttering in detector materials.
}
\label{AngleMatching}
\end{figure}

Angle information from the new detectors enables further reduction of low \pT~muons in addition to the position matching algorithm.
The $d\theta$ information from the new detectors is defined by the angular difference between the angle of the reconstructed track and the angle of the straight line connecting the nominal interaction point and the track position.
The $d\theta$ of the track should have a value near zero when the muon is produced at the detector center and enters the detectors straight.
However, the interaction point can differ from the detector center within the beam spot size in the z-direction of approximately 10 cm.
In addition, when multiple scattering occurs inside detector materials, especially in the calorimeter, there is a shift in their direction from their initial path.
Low \pT~muons imitating high \pT~muons with these two reasons cannot be reduced by applying position matching as shown in \figurename~\ref{AngleMatching}.
In this case, the $d\theta$ from the new detectors is different for low \pT~and high \pT~muons even with the same position difference.
By combining the $d\theta$ information with the position difference, further reduction of low \pT~muons is achieved. 

\subsection{Hardware design of Sector Logic}
\begin{figure}[tbh]
  \centerline{\includegraphics[width=3.5in]{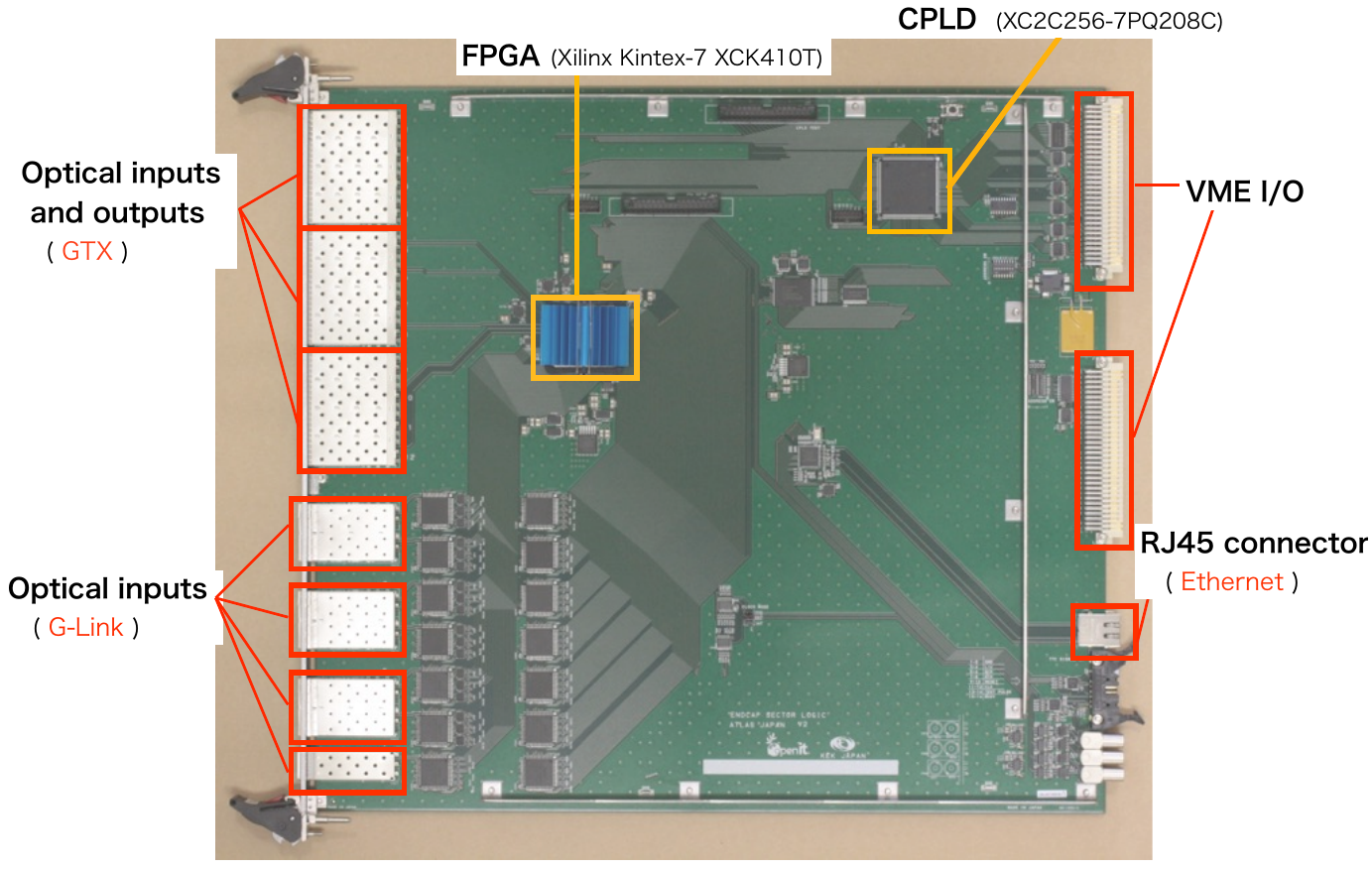}}
  \caption{Block diagram of the endcap SL in Run3. Xilinx Kintex-7 FPGA is implemented on a VME 9U board as the main processor.}
  \label{Run3_SL}
\end{figure}

Figure~\ref{Run3_SL} shows the hardware design of the SL board in \mbox{Run 3}.
In order to handle data from various detectors, the endcap SL is required to have enough I/O ports.
13 G-Link \cite{G-Link} connections, which require large amount of I/O ports, are used to receive data from the TGC-BW and TileCal.
273 I/Os are required to establish G-Link connection because 21 I/Os are used by G-Link per channel. 
Data from the NSW, RPCBIS78 and TGC EI will be received by GTX \cite{GTX} transceivers.
GTX is a multi-gigabit transceiver for Xilinx Kintex-7 FPGAs, supporting line rates up to 12.5 Gbps.
Large amounts of  memory resources are also required to implement new coincidence logic with new detectors.

Xilinx Kintex-7 FPGA \cite{Kintex-7} (XC7K410T-1FFG900) is selected as the main processor of the endcap SL which meets these requirements.
XC7K410T has 500 I/O pins and 16 GTX \cite{GTX} transceivers which is enough to handle data from all detectors.
XC7K410T has 795 Block RAMs (BRAMs) \cite{BRAM}, which is about 20 times the memory resource compared to the FPGA used in \mbox{Run 2}.
BRAM is a RAM module which provides storage for large set of data up to 36 Kb.
LUTs for coincidence logic will be implemented on BRAMs.
Thus, larger amount of BRAM resources leads to improved trigger performance.

Complex Programmable Logic Device (CPLD) is also placed on the SL board to control the VME bus.
Non-volatile memory on the CPLD enables the configuration of FPGA at power up.
FPGA configuration by Byte Peripheral Interface (BPI) memory, containing data of the firmware design, is also controlled by CPLD.

\subsection{Firmware implementation for the new coincidence logic}
The new coincidence logic will be implemented on the FPGA with fixed latency.
The SL is required to send trigger candidates at 53 LHC clocks after the bunch crossing.
Considering the arrival time of the track information from the NSW and serialization, data transfer and trigger logic before sending information to the subsequent boards, the new coincidence logic is required to be finished within 2 LHC clocks.

One track candidate from the TGC-BW will be compared with several track candidates from the new detectors.
For example, the NSW will send 16 track candidates to the endcap SL at the maximum.
Calculating the \pT~using the NSW track candidates in parallel by placing 16 identical LUTs is the simplest implementation to achieve short latency.
However, this implementation makes the memory usage 16 times as large.
To minimize the memory usage, processing the NSW track candidates in serial would be an alternative implementation.
Still the latency would be 16 times longer which would not meet the requirements.

\begin{figure}[tbh]
\centerline{\includegraphics[width=3.5in]{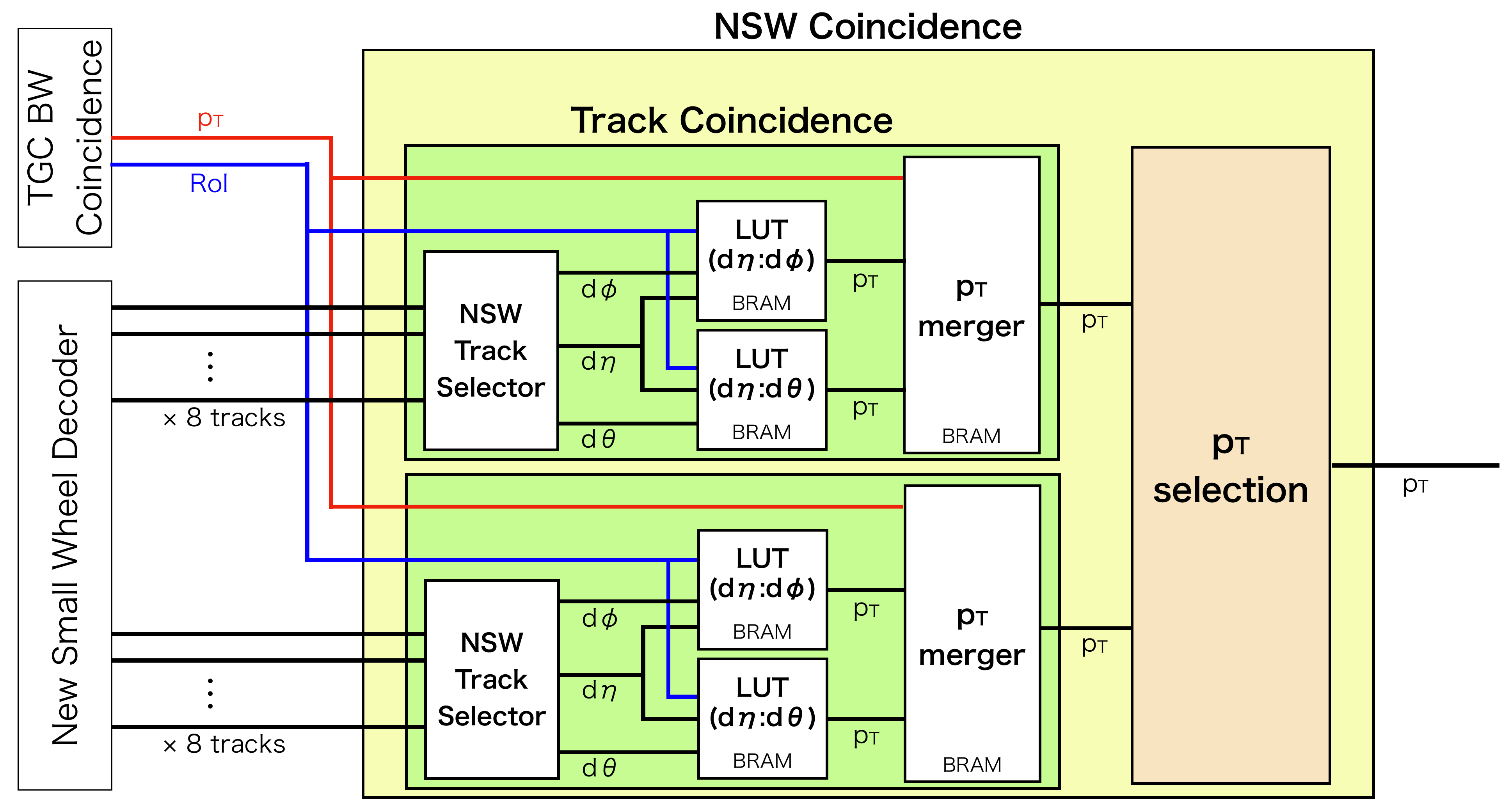}}
\caption{
  Block diagram for the new coincidence logic.
  The Track Coincidence module and \pT~Selection module operate at 320 MHz clock to calculate the \pT~for 16 track candidates in 2 LHC clocks.
}
\label{Run3_Firmware}
\end{figure}

To achieve short latency and minimization of memory usage, the firmware is designed as in \figurename~\ref{Run3_Firmware}.
This firmware design consists of two modules operated at 320 MHz clock.

\vspace{\baselineskip}
\subsubsection{Track Coincidence}
Two Track Coincidence modules are placed in parallel with identical LUTs and receive eight NSW track candidates per-module.
The NSW Track Selector receives eight NSW track candidates and sends track information to the LUTs for position and angle matching in serial.
16 tracks can be processed in one LHC clock by two Track Coincidence modules using the 320 MHz clock, which is eight times faster than the LHC clock.
After the \pT~extraction from LUTs is finished for each track, the \pT~Merger module is used to choose which \pT~to take as the final decision, which is basically the highest \pT.

\vspace{\baselineskip}
\subsubsection{\pT~Selection}
The \pT~Selection module receives 2 candidates per clock tick.
To select the highest \pT~among the 16 candidates, the \pT~Selection module selects the highest \pT~among 3 candidates per clock tick, the 2 candidates from the Track Coincidence module and the candidate with the highest \pT~selected so far.
A register is placed in the \pT~Selection module to keep the candidate selected so far.
After receiving 16 candidates, the candidate kept in the register is sent back to the LHC clock domain.

%\subsection{Performance of the new coincidence logic firmware}

\subsection{Performance of Level-1 endcap muon trigger}
\begin{figure}[tbh]
\centerline{\includegraphics[width=3.0in]{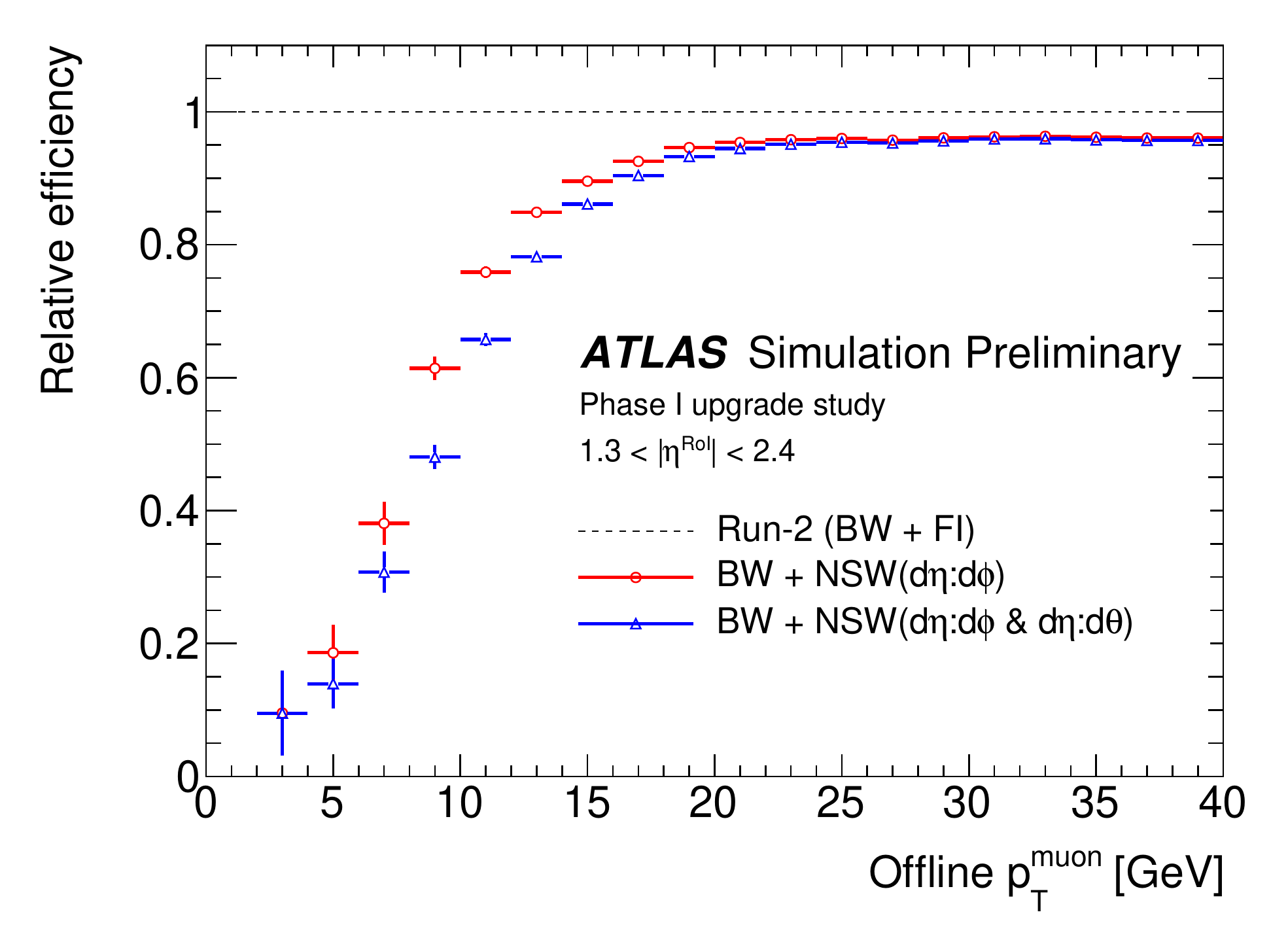}}
\caption{
  The \pT~dependency of the trigger efficiency relative to the \mbox{Run 2} trigger \cite{L1Public}.
  The red (blue) line indicates the relative trigger efficiency by including the position (position and angle) matching algorithm.
}
\label{Run3_Efficiency}
\end{figure}

Rejection power for low \pT~muons is estimated from a single muon MC simulation sample.
Track reconstruction efficiency of the NSW is assumed to be 97\% and included in the calculation.
Figure~\ref{Run3_Efficiency} shows the \pT~dependency of the relative trigger efficiency compared to the \mbox{Run 2} trigger efficiency.
Higher reduction for low \pT~muons relative to \mbox{Run 2} trigger is seen by including the position and angle matching algorithm.

\begin{figure}[tbh]
\centerline{\includegraphics[width=3.0in]{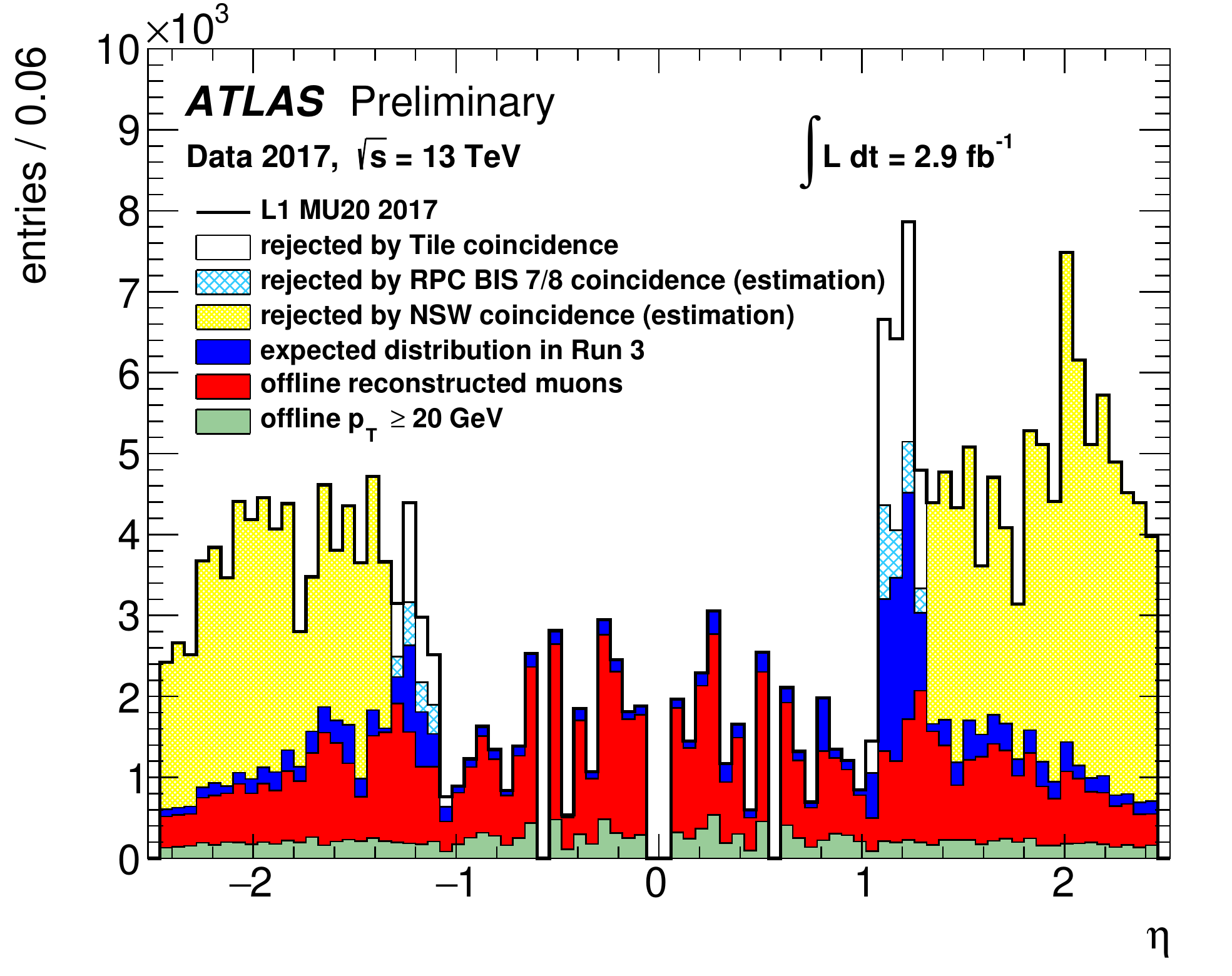}}
\caption{
  The $\eta$ distribution of trigger candidates triggered by L1 MU20 expected in \mbox{Run 3}\cite{L1Public}.
  The yellow and blue hatching area shows the number of reduced low \pT~and fake muons by introducing the new coincidence logic by the NSW and RPCBIS78, respectively.
}
\label{Run3_Rate}
\end{figure}

Rejection power for the fake muons is estimated from 2017 data since the fake muons cannot be modeled by MC simulations.
Figure~\ref{Run3_Rate} shows the $\eta$ distribution of trigger candidates triggered by the L1 MU20 trigger expected in \mbox{Run 3}.
90\% of the fake muons are reduced by new inner muon detectors compared to \mbox{Run 2} logic.
The expected trigger rate in \mbox{Run 3} is 13 kHz, which meets the requirements for \mbox{Run 3}.

%%%%%%%%%%%%%%%%%%%%%%%%%%%%%%%%%%%%%%%%%%%%%%%
%% HL-LHC                                    %%
%%%%%%%%%%%%%%%%%%%%%%%%%%%%%%%%%%%%%%%%%%%%%%%
\section{PHASE-II UPGRADE OF \\ THE ATLAS LEVEL-0 MUON TRIGGER}

\begin{figure}[tbh]
\centerline{\includegraphics[width=3.5in]{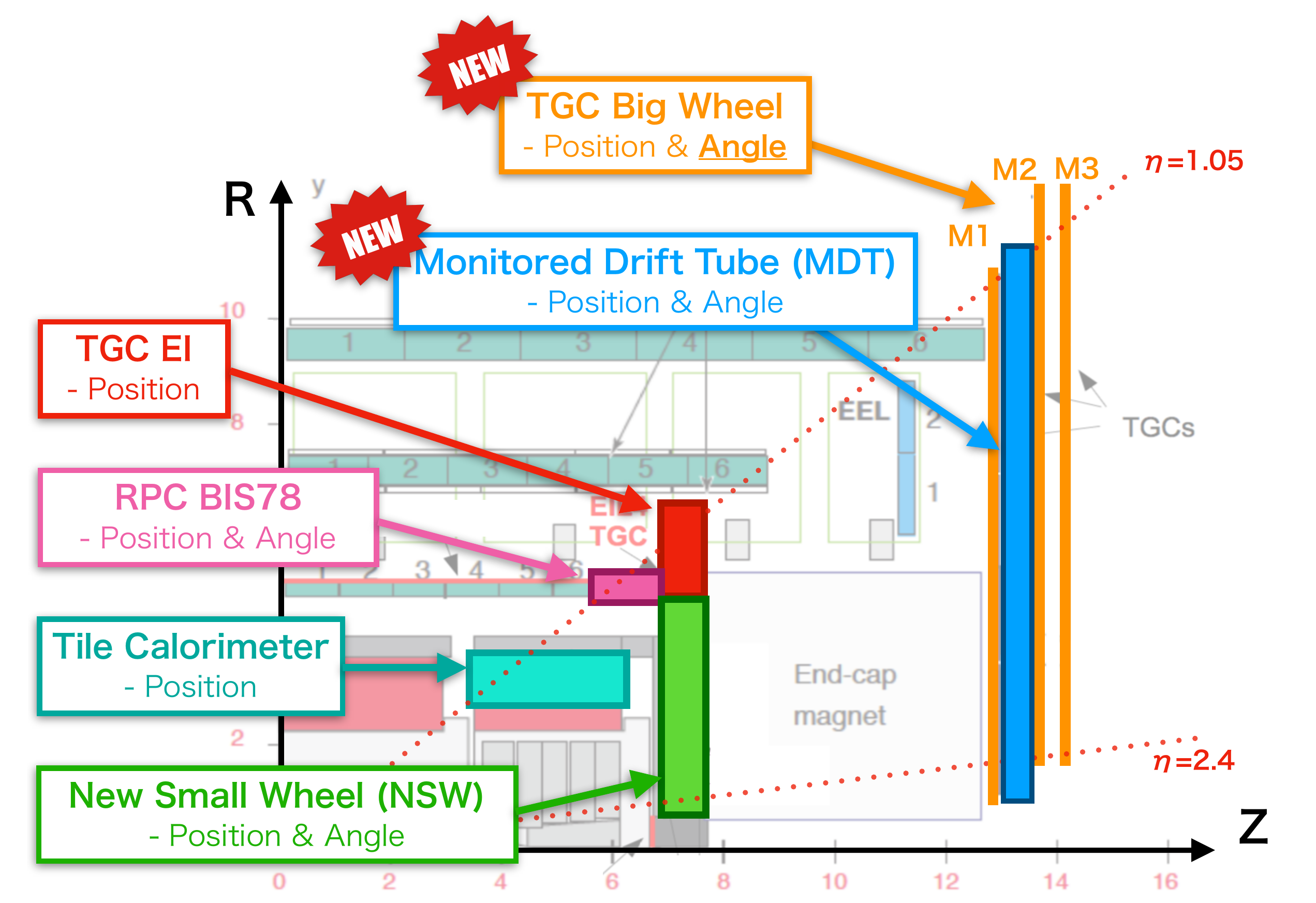}}
\caption{Layout of the ATLAS muon spectrometer in the \mbox{HL-LHC}. Hardware-based track reconstruction will be implemented to obtain position and angle information with the TGC-BW. MDTs used in the software-based trigger for the current trigger system  will be moved to the hardware-based trigger for precise \pT~determination.}
\label{HLLHC_Detector}
\end{figure}

Figure~\ref{HLLHC_Detector} shows detectors used in the L0 endcap muon trigger for the \mbox{HL-LHC}.
In the new endcap muon trigger system, the sector logic will receive all TGC hit information from the new boards on the detector side.
Hardware-based track reconstruction using all TGC hit information will be enabled to measure \pT~of track candidates with higher resolution.
During \mbox{Run 2}, Monitored Drift Tubes (MDTs) were used for precise segment reconstruction only in the software-based trigger due to their long latency.
MDTs will remain in the software-based trigger after the Phase-I Upgrade since L1 trigger latency will not be extended.
Due to the extension of the L0 trigger latency from 2.5 \ms~to 10 \ms~during the Phase-II Upgrade, MDTs will be used in the hardware-based trigger for precise \pT~determination. 

\subsection{TGC track reconstruction}
\begin{figure}[tbh]
  \centerline{\includegraphics[width=3.5in]{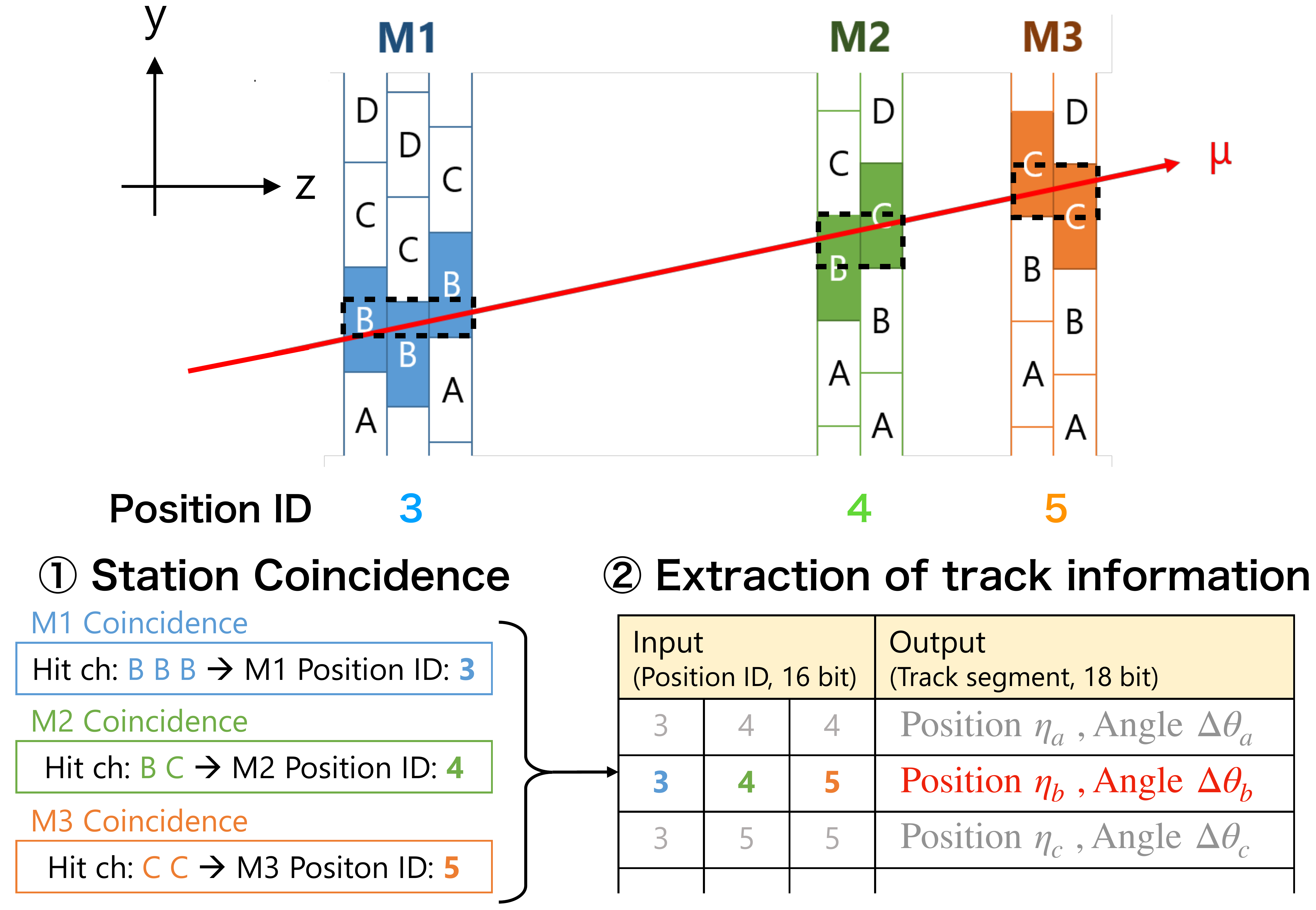}}
  \caption{
    Schematic drawing of the pattern matching algorithm.
    TGC tracks are extracted in two steps.
    First, position ID is defined by taking coincidence in each station.
    Second, hit patterns corresponding to the combination of the position ID is extracted from the pattern list. 
  }
  \label{PatternMatching}
\end{figure}

Tracks will be reconstructed with a pattern matching algorithm using all hits from the TGC-BW.
The TGC-BW consists of three stations, M1, M2 and M3, which consist of three, two and two layers respectively.
Figure~\ref{PatternMatching} shows the main concept of the pattern matching algorithm.
After receiving all TGC-BW hit information, coincidence is taken to define a position in each station.
Combination of the position in each station is compared with a predefined list of hit patterns, which has position and angle information for high \pT~muons.

In the current trigger scheme, at least two (one) hits in the M1 station and at least three (two) hits in the M2 and M3 stations are required for wires (strips) on the front-end boards.
Due to this requirement, the muon efficiency of TGC wires and strips is limited to 95.6\% and 96.9\% respectively.
Combining these two efficiencies, the muon efficiency of the TGC-BW is estimated as 91.8\% in the current system. 
The muon efficiency of TGC wires and strips for each layer are assumed to be 92.7\% and 92.1\% respectively and included in the calculation.
In the \mbox{HL-LHC} trigger scheme, at least five (four) hits in the M1, M2 and M3 stations are required.
The muon efficiency of TGC wires and strips is improved to 98.2\% and 97.8\% respectively with looser coincidence.
Combining these two efficiencies, the muon efficiency of the TGC-BW is estimated as 96.0\%.
The new trigger scheme is expected to improve the trigger efficiency by 4.2\% compared to the current trigger scheme.  

%The angle resolution of the pattern matching algorithm is estimated by comparing the polar angle of tracks reconstructed by the pattern matching algorithm and ATLAS full offline analysis.
%As shown in \figurename~\ref{AngleResolution}, the average angle resolution is estimated as about 4 mrad from a single muon MC simulation sample.

%\begin{figure}[tbh]
%  \centerline{\includegraphics[width=3.0in]{fig/AngleResolution.pdf}}
%  \caption{
%    Distributions of polar angle difference ($\Delta\theta$) between track segments reconstructed in TGC-BW and MDT \cite{Phase2_TDR}.
%    Red, blue and green histograms show the $\Delta\theta$ distribution when seven, six and five layers have matched respectively.
%    The total of the three histograms are shown in the black histogram.
%  }
%  \label{AngleResolution}
%\end{figure}

\subsection{Hardware design of Sector Logic}
\begin{figure}[tbh]
  \centerline{\includegraphics[width=3.5in]{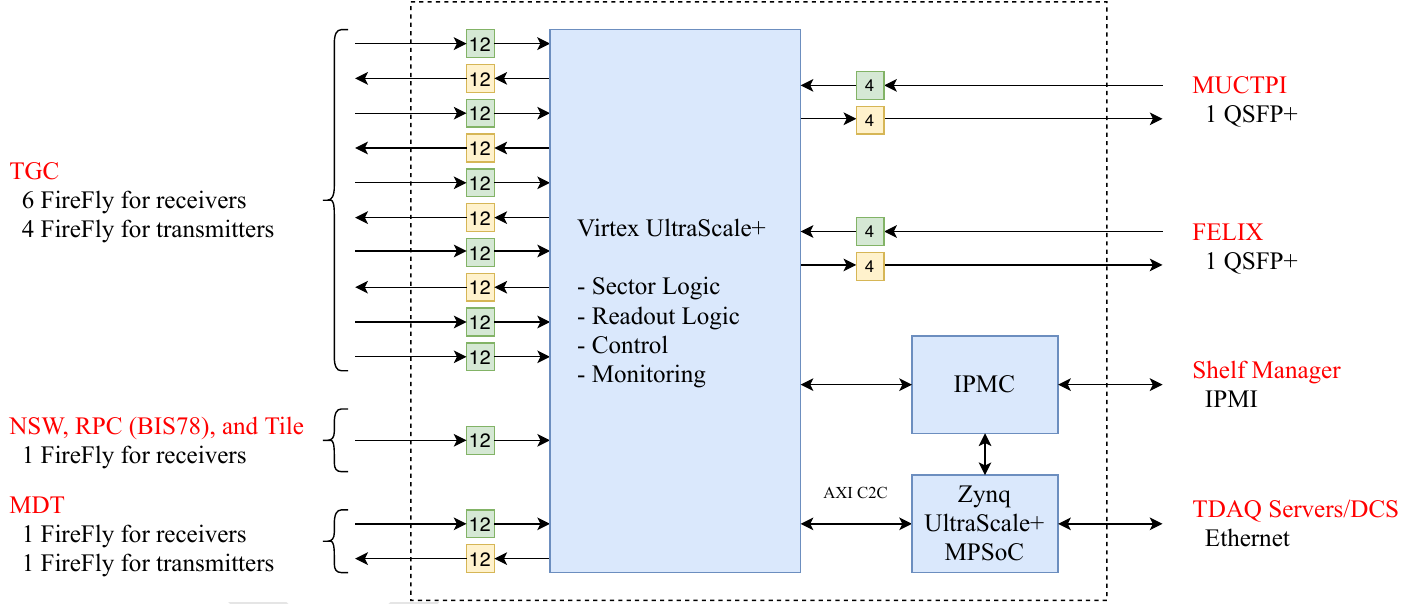}}
  \caption{Block diagram of the endcap SL in the \mbox{HL-LHC}. Xilinx Virtex UltraScale+ FPGA will be implemented on a ATCA blade as the main processor.}
  \label{HLLHC_SL}
\end{figure}

Figure~\ref{HLLHC_SL} shows the hardware design of the SL board in the \mbox{HL-LHC}.
In order to handle data from every TGC channel, the endcap SL is required to have enough I/O ports.
A few hundred Mbits of memory resources are also required to implement pattern lists for track reconstruction.
Xilinx Virtex UltraScale+ FPGA \cite{VirtexUltraScale+} (XCVU9P-1FLGA2577E) is selected as the main processor of the endcap SL which meets these requirements.
XCVU9P has 448 I/O pins and 120 GTY \cite{GTY} transceivers which is enough to handle data from all detectors.
GTY is a multi-gigabit transceiver for Xilinx UltraScale FPGAs, supporting line rates up to 32.75 Gbps in UltraScale+ FPGAs.
XCVU9P has 2160 BRAMs and 960 UltraRAMs \cite{UltraRAM}, which is about ten times the memory resource compared to the FPGA used in \mbox{Run 3}.
UltraRAM is a new memory block with large capacity up to 288 Kb implemented in UltraScale+ families.

FireFly \cite{FireFly} and QSFP+ modules will be placed on the SL board to manage data transfer with optical connections.
FireFly modules are capable of handling 12 channels up to 16 Gbps per channel.
QSFP+ modules are capable of handling four channels up to 10 Gbps per channel.

CERN-developed Intelligent Platform Management Controller (IPMC) for ATCA blades will be implemented for control and configuration at power up through the ATCA shelf manager. 

Multiprocessor System-on-Chip (MPSoC) \cite{MPSoC} device implemented on the SL board will be the interface for the central ATLAS Run Control, Configuration, and Monitoring of the status registers of the FPGA.

\subsection{Firmware implementation for the TGC track reconstruction}
The TGC track reconstruction logic will be implemented on the FPGA with fixed latency.
Arrival of the TGC hit signals to the endcap SL is estimated to be 0.888 \ms~after the bunch crossing.
The estimated latency of the TGC track reconstruction is 1.013 \ms~after the bunch crossing.
Thus, the firmware design is required to reconstruct tracks in 0.125 \ms.

\begin{figure}[tbh]
  \centerline{\includegraphics[width=3.5in]{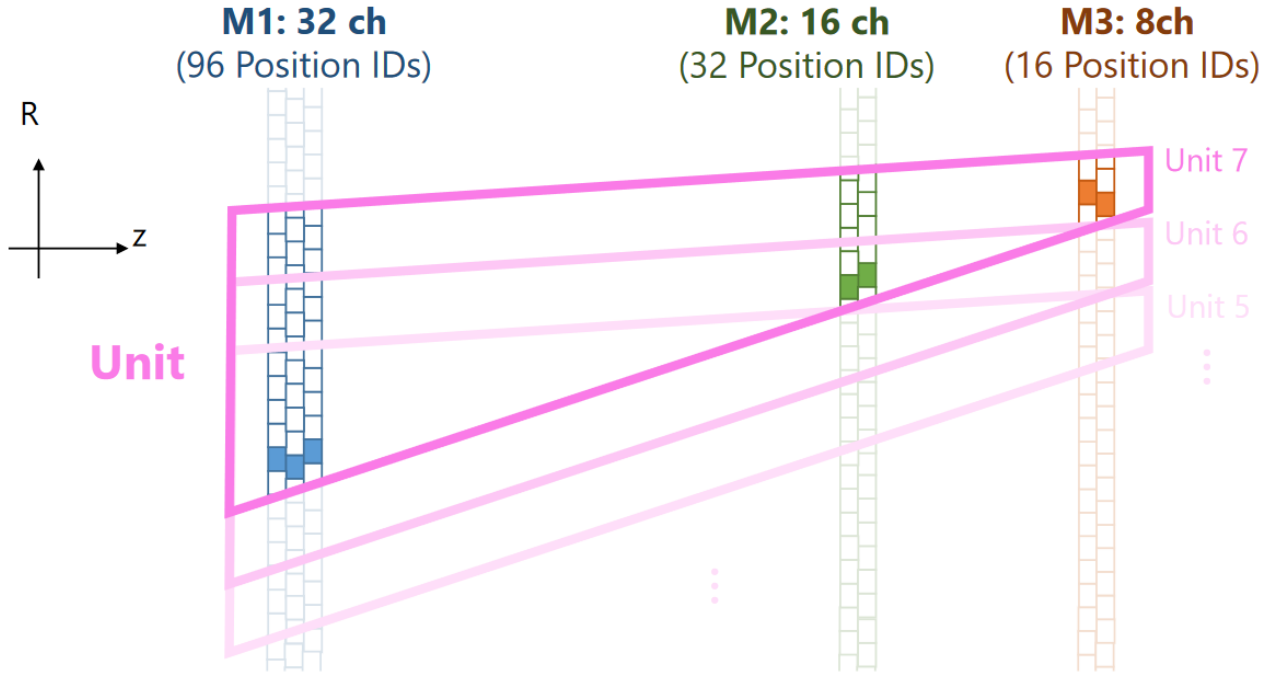}}
  \caption{
    Definition of the subdivided region (``Unit'') of the TGC-BW.
  }
  \label{Unit}
\end{figure}

This algorithm is processed parallelly in subdivided small regions (``Unit'') to reduce redundant pattern lists as shown in \figurename~\ref{Unit}.
Units are defined to include TGC hits from muons (both $\mu^{+}$ and $\mu^{-}$) with \pT~as low as 4 GeV.
This leads to a triangular-shaped region which consists of 8 wire channels per layer in M3, 16 wire channels per layer in M2 and 32 wire channels per layer in M1.
These Units are subdivided into four regions (``Subunit'') and one URAM block is allocated for each Subunit to store pattern lists.
A Subunit is defined by dividing the 8 wire channels consisting a Unit into four regions, two wire channels per layer in M3.

\begin{figure}[tbh]
\centerline{\includegraphics[width=3.5in]{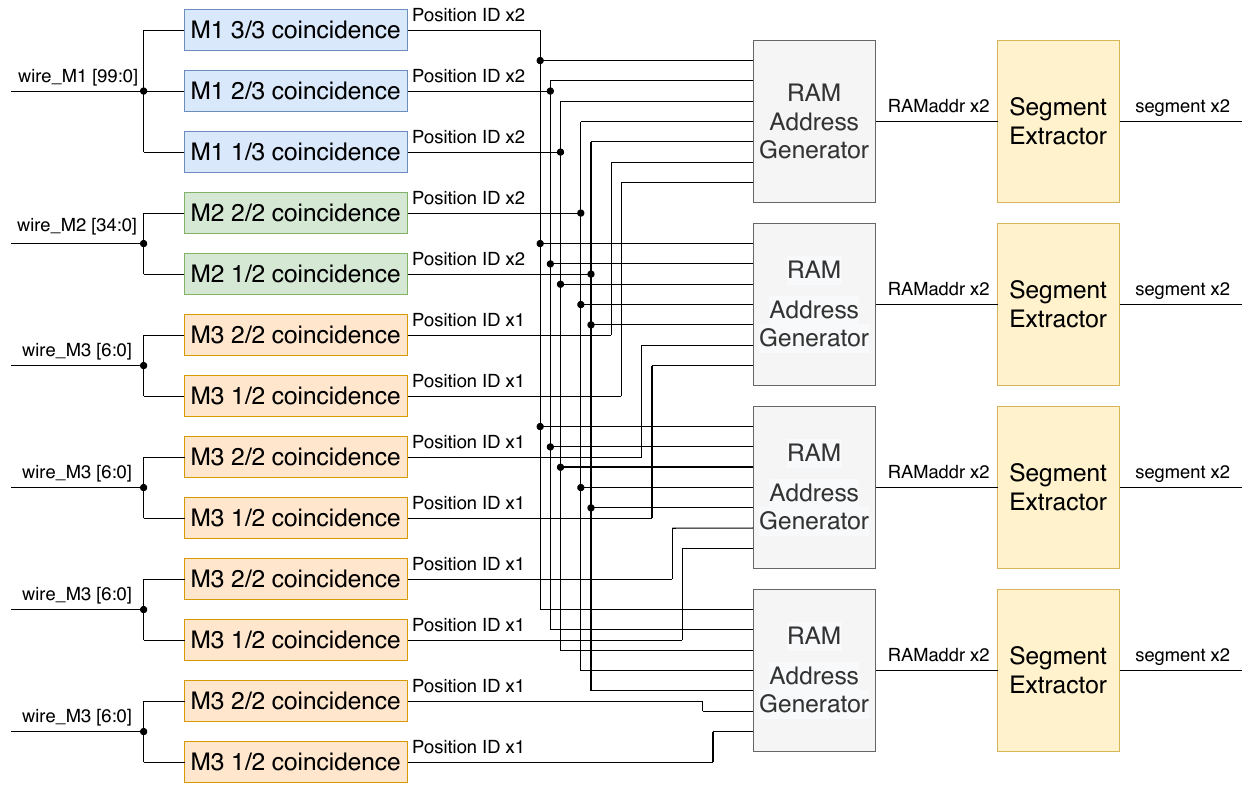}}
\caption{
  Block diagram for the TGC track reconstruction.
}
\label{HLLHC_Firmware}
\end{figure}

A firmware block diagram for the TGC track reconstruction is shown in \figurename~\ref{HLLHC_Firmware}.
After receiving hit information from the TGC-BW for every bunch crossing at 40 MHz, a 160 MHz clock synchronous to the 40 MHz LHC clock is used to process the hits. 
The firmware block for TGC track reconstruction consists of three modules.

\vspace{\baselineskip}
\subsubsection{Station Coincidence}
Coincidence of the TGC hits are taken in each station of the TGC-BW and coincidence results (``Position IDs'') are output for each station.
There are seven types of modules for station coincidence, corresponding to the number of hits required in each station.
For example, the ``M1 2/3 coincidence'' block requires hits in two layers with no hit in the remaining one layer in the M1 station.
Position IDs from the center of the Unit region output by M1 and M2 station coincidence modules are preferentially sent to the subsequent modules. 
For M3 station coincidence modules, Position IDs with smaller $\eta$ position are prioritized.
This prioritization scheme is optimized to select track candidates with higher \pT.
M1 and M2 station coincidence modules output two Position IDs and M3 station coincidence modules output one Position ID.

\vspace{\baselineskip}
\subsubsection{RAM Address Generator}
The RAM Address Generator module receives Position IDs from the Station Coincidence module and combines them to create a RAM address for the corresponding hit pattern.
The RAM address is described in 12 bits using M1, M2 and M3 Position IDs, which are described in 5, 5 and 2 bits respectively.
The considered patterns for the combination of Position IDs are shown in Table~\ref{Patterns}.
Combination of Position IDs with larger number of hits are prioritized and a maximum of eight hit pattern candidates are obtained per Subunit.
When the candidates have the same coincidence pattern, candidates with smaller $\eta$ are preferentially selected.

\begin{table}
\centering
  \caption{
    Definition of coincidence patterns
  }
  \label{Patterns}
  \setlength{\tabcolsep}{3pt}
  \begin{tabular}{|c|ccc|c|}
    \hline
    & \multicolumn{3}{c|}{Hit pattern} & \\ \hline
    Coincidence pattern & M1 & M2 & M3 & Fraction \\ \hline \hline
    7/7 & 3/3 & 2/2 & 2/2 & 0.649 \\ \hline
    6/7A & 2/3 & 2/2 & 2/2 & 0.124 \\ \hline
    6/7B & 3/3 & 1/2 & 2/2 & 0.083 \\ \hline
    6/7C & 3/3 & 2/2 & 1/2 & 0.083 \\ \hline
    5/7A & 2/3 & 1/2 & 2/2 & 0.016 \\ \hline
    5/7B & 2/3 & 2/2 & 1/2 & 0.016 \\ \hline
    5/7C & 3/3 & 1/2 & 1/2 & 0.011 \\ \hline
    5/7D & 1/3 & 2/2 & 2/2 & 0.008 \\ \hline \hline
    Total & & & & 0.988 \\ \hline
    %\multicolumn{5}{p{240pt}}{
    \multicolumn{5}{p{220pt}}{	
    \footnotesize{The coincidence patterns for TGC wire track reconstruction are listed in descending order of the fraction.
    The fraction of muons for each coincidence pattern is calculated by assuming that the hit efficiency for each TGC layer is 94\%.}
    }\\
  \end{tabular}
\end{table}
   
\vspace{\baselineskip}
\subsubsection{Segment Extractor}
The Segment Extractor module receives two RAM addresses per clock, using four clock ticks in total to receive eight RAM addresses per Subunit.
URAM blocks are configured in the True Dual Port mode, and track information corresponding to the RAM address are extracted from the RAM per clock tick.

%\subsection{Performance of the TGC track reconstruction firmware}
%Initial test of TGC track reconstruction for one Unit of the forward trigger sector ($2.16 < \eta < 2.19, 0.26 < \phi < 0.52$) was performed with Xilinx Virtex UltraScale+ FPGA VCU118 evaluation kit\cite{EvaluationBoard}.
%XCVU9P FPGA is implemented on the VCU118 evaluation kit.
%Input data was generated by extracting TGC hit information from 1000 muons with \pT~= \mbox{20 GeV} using a single muon MC sample. 
%Only events with one hit per-layer are used as the input data.
%All tracks were successfully reconstructed and showed angle resolution of 2.4 mrad as shown in \figurename~\ref{InitialTest}.

%\begin{figure}[tbh]
%\centerline{\includegraphics[width=3.0in]{fig/InitialTest.pdf}}
%\caption{
%  Distribution of polar angle difference ($\Delta\theta$) between tracks reconstructed in TGC track reconstruction firmware and true track\cite{L0Public}.
%}
%\label{InitialTest}
%\end{figure}

\subsection{Performance of Level-0 endcap muon trigger}
\begin{figure}[tbh]
\centerline{\includegraphics[width=3.0in]{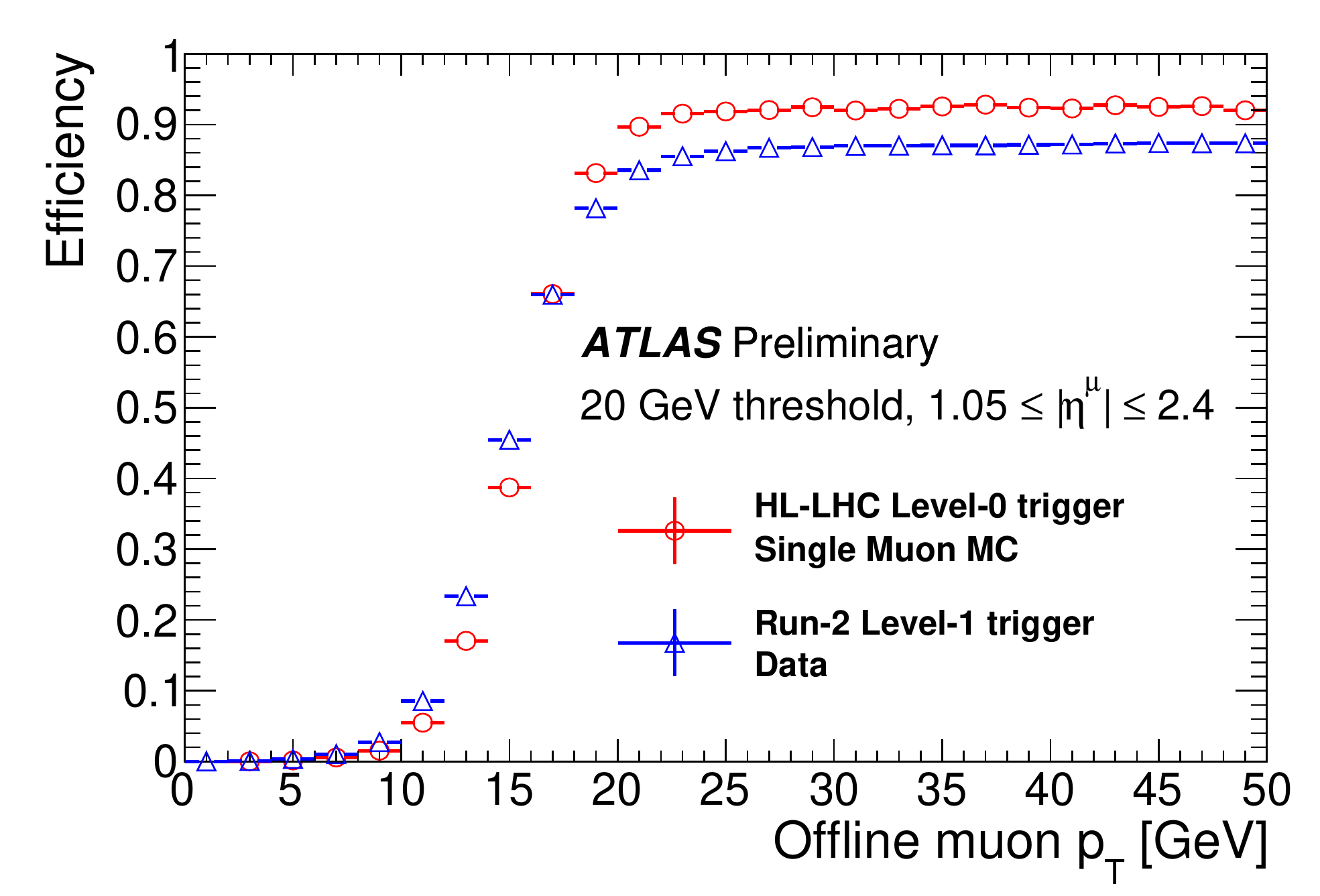}}
\caption{
  The \pT~dependency of trigger efficiency for the L0 muon trigger with the \mbox{HL-LHC} scheme (red) and with the \mbox{Run 2} scheme (blue)\cite{L0Public}.
}
\label{HLLHC_Efficiency}
\end{figure}

Trigger performance of the L0 endcap muon trigger is evaluated.
However, the precise information from the MDTs is not used in the evaluation.
Figure~\ref{HLLHC_Efficiency} shows the expected efficiency of the new trigger algorithm with respect to offline muons in a single muon  MC simulation sample.
Compared to the \mbox{Run 2} trigger scheme, higher efficiency in the plateau region and better rejection for the low \pT~muons are obtained due to the looser coincidence and the improved angular resolution, respectively.
 
\begin{figure}[tbh]
\centerline{\includegraphics[width=3.0in]{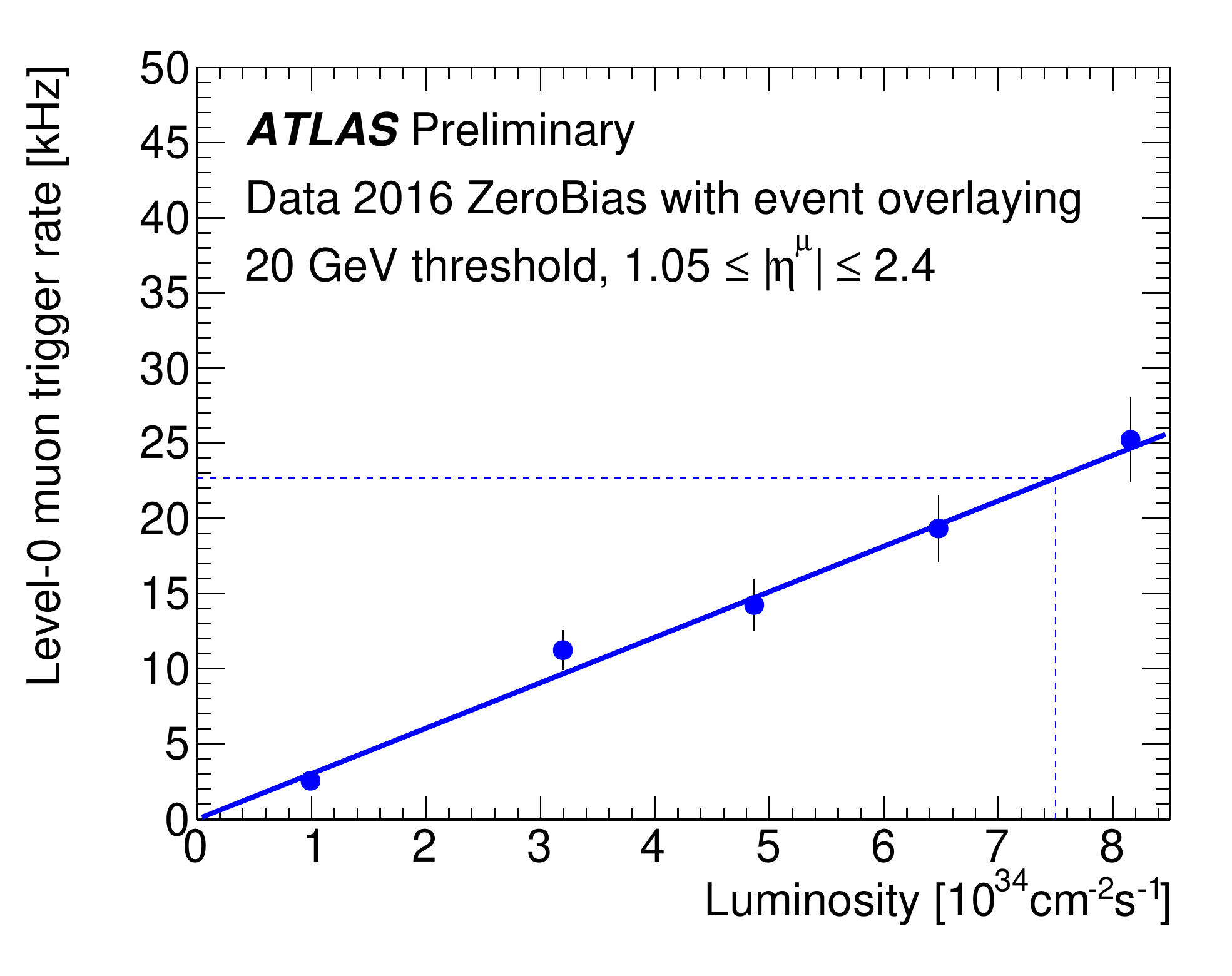}}
\caption{
  Luminosity dependence of the estimated trigger rate of the L0 single muon trigger at \pT~threshold of \mbox{20 GeV} \cite{L0Public}.
  The lowest luminosity point corresponds to the \mbox{Run 2} data taken with a random trigger and the higher luminosity points are produced by overlaying the \mbox{Run 2} events.
}
\label{HLLHC_Rate}
\end{figure}

Figure~\ref{HLLHC_Rate} shows the estimated trigger rate from \mbox{Run 2} data taken with a random trigger to reproduce higher luminosity expected in the \mbox{HL-LHC}.
The trigger rate for a \mbox{20 GeV} threshold is about 23 kHz, which constitutes only about 2.3\% of the assumed total L0 trigger rate of 1 MHz.
Further rate reduction is expected by using the MDTs which improves the \pT~resolution of the track candidates.

\section{Conclusion}
Continuous upgrades of the hardware-based (Level-1, -0) endcap muon trigger is planned for \mbox{Run 3} and the \mbox{HL-LHC} to keep the physics acceptance.
For \mbox{Run 3}, new detectors with finer granularity track information will be installed inside the toroidal magnetic field.
New coincidence logic using position and angle information of the new detectors was suggested.
The estimated L1 endcap muon trigger rate for a \mbox{20 GeV} threshold is about 13 kHz at an instantaneous luminosity of \mbox{2.0 \lumi,} which meets the requirements for \mbox{Run 3}.
The new trigger processor board (SL) has been produced for \mbox{Run 3} to handle data from various detectors and implement the new coincidence logic which requires large amounts of resources.
For the \mbox{HL-LHC}, the trigger and readout electronics will be replaced to extend the L0 trigger rate and latency.
Using full-granularity information from the TGC-BW, fast track segment reconstruction will be implemented to the SL board in \mbox{Run 3}.
The new L0 endcap muon trigger scheme shows about 4\% higher efficiency compared to the current trigger system.
In addition, the estimated trigger rate for a \mbox{20 GeV} threshold is about 23 kHz at an instantaneous luminosity of \mbox{7.5 \lumi}, which constitutes only about 2.3\% of the assumed total L0 trigger rate.
Further rate reduction is expected by adding the MDTs in the L0 trigger.
Design of the SL board in the \mbox{HL-LHC} has been introduced to implement the TGC track reconstruction algorithm.
%The initial test of the TGC track reconstruction was achieved by using an evaluation board and showed expected angle resolution and reconstruction efficiency.

\section*{References}

\def\refname{\vadjust{\vspace*{-1em}}} %Please don't do this in a real paper.


\begin{thebibliography}{00}
\bibitem{LHC} L. Evans and P. Bryant, ``LHC Machine,'' \textit{JINST}, vol. 3, p. S08001, 2008.
\bibitem{schedule} ``HL-LHC : Project Schedule'' Jan. 12, 2012. [Online]. Available: \url{https://project-hl-lhc-industry.web.cern.ch/content/project-schedule}
\bibitem{ATLAS} ATLAS Collaboration, ``The ATLAS Experiment at the CERN Large Hadron Collider,'' \textit{JINST}, vol. 3, p. S08003, 2008.
\bibitem{Phase1TDR} ATLAS Collaboration, ``Technical Design Report for the Phase-I Upgrade of the ATLAS TDAQ System,'' 2013. [Online]. Available: \url{https://cds.cern.ch/record/1602235}
\bibitem{NSW} ATLAS Collaboration, ``New Small Wheel Technical Design Report,'' 2013. [Online]. Available: \url{https://cds.cern.ch/record/1552862}
\bibitem{L1Public} ATLAS Collaboration, ``L1 Muon Trigger Public Results.'' [Online]. Available: \url{https://twiki.cern.ch/twiki/bin/view/AtlasPublic/L1MuonTriggerPublicResults}
\bibitem{TGC} ATLAS Collaboration, ``ATLAS muon spectrometer : Technical Design Report,'' 1997. [Online]. Available: \url{https://cds.cern.ch/record/331068}
\bibitem{RPCBIS78} ATLAS Muon Collaboration, ``Proposal of upgrade of the ATLAS muon trigger in the Barrel - End Cap transition region with RPCs,'' in \textit{PoS TIPP2014}, 2014, p. 117.
\bibitem{TileCal} ATLAS Collaboration, ``ATLAS tile calorimeter : Technical Design Report,'' 1996. [Online]. Available: \url{https://cds.cern.ch/record/331062}
\bibitem{Micromegas} Y. Giomataris, P. Rebourgeard, J. Robert, and G. Charpak, ``MICROMEGAS : a high-granularity position-sensitive gaseous detector for high particle-flux environments,'' \textit{Nucl.Instrum.Meth.}, vol. A376, p. 29-35, 1996.
\bibitem{G-Link} ``Agilent HDMP-1032/1034 transmitter/receiver chip-set data sheet,'' Agilent Technologies. [Online]. Available: \url{https://www.asc.ohio-state.edu/physics/cms/cfeb/datasheets/hdmp1032.pdf}  
\bibitem{GTX} ``7 Series FPGAs GTX/GTH Transceivers User Guide'' Xilinx Inc. [Online]. Available: \url{https://www.xilinx.com/support/documentation/user_guides/ug476_7Series_Transceivers.pdf}
\bibitem{Kintex-7} ``7 Series FPGAs Data Sheet: Overview,'' Xilinx Inc. [Online]. Available: \url{https://www.xilinx.com/support/documentation/data_sheets/ds180_7Series_Overview.pdf}
\bibitem{BRAM} ``7 Series FPGAs Memory Resources,'' Xilinx Inc. [Online]. Available: \url{https://www.xilinx.com/support/documentation/user_guides/ug473_7Series_Memory_Resources.pdf}
%\bibitem{Phase2_TDR} ATLAS Collaboration ``Technical Design Report for the Phase-II Upgrade of the ATLAS TDAQ System'' Geneva, Sep. 2017. [Online]. Available: \url{https://cds.cern.ch/record/2285584}
\bibitem{VirtexUltraScale+} ``UltraScale Architecture and Product Data Sheet: Overview'' Xilinx Inc. [Online]. Available: \url{https://www.xilinx.com/support/documentation/data_sheets/ds890-ultrascale-overview.pdf}  
\bibitem{GTY} ``UltraScale Architecture GTY Transceivers User Guide'' Xilinx Inc. [Online]. Available: \url{https://www.xilinx.com/support/documentation/user_guides/ug578-ultrascale-gty-transceivers.pdf}
\bibitem{UltraRAM} ``UltraRAM: Breakthrough Embedded Memory Integration on UltraScale+ Devices'' Xilinx Inc. [Online]. Available: \url{https://www.xilinx.com/support/documentation/white_papers/wp477-ultraram.pdf}
\bibitem{FireFly} ``FIREFLY : APPLICATION DESIGN GUIDE'' Samtec Inc. [Online]. Available: \url{http://suddendocs.samtec.com/ebrochures/firefly-brochure.pdf}.
\bibitem{MPSoC} ``Zynq UltraScale+ MPSoC Data Sheet: Overview'' Xilinx Inc. [Online]. Available: \url{https://www.xilinx.com/support/documentation/data_sheets/ds891-zynq-ultrascale-plus-overview.pdf}
%\bibitem{EvaluationBoard} ``VCU118 Evaluation Board User Guide'' Xilinx Inc. [Online]. Available: \url{https://www.xilinx.com/support/documentation/boards_and_kits/vcu118/ug1224-vcu118-eval-bd.pdf}
\bibitem{L0Public} ATLAS Collaboration, ``Level-0 TGC Trigger Performance of trigger algorithms in software and firmware implementations'' [Online]. Available: \url{https://twiki.cern.ch/twiki/bin/view/AtlasPublic/L0MuonTriggerPublicResults}
\end{thebibliography}
\end{document}